\documentclass[conference]{IEEEtran}
\usepackage{graphicx}
\usepackage{float}
\usepackage{cite}
\usepackage{hyperref}
\usepackage{multicol, blindtext}
\ifCLASSINFOpdf
\else
\fi
\usepackage{booktabs}

\begin{document}
%
\title{SEDAT:\underline{S}ecurity \underline{E}nhanced \underline{D}evice \underline{A}ttestation \\ with \underline{T}PM2.0}

\author{\IEEEauthorblockN{Avani Dave, Monty Wiseman David Safford}
	\IEEEauthorblockA{Department of Computer Science and Electrical Engineering, University of Maryland, Baltimore County}
}


%

\maketitle

\begin{abstract} Remote attestation is one of the ways to verify the state of an untrusted device. Earlier research has attempted remote verification of a device's state using hardware, software, or hybrid approaches. Majority of them have used Attestation Key as a hardware root of trust, which does not detect hardware modification or couterfiet issues. In addition, they do not have a secure communication channel between verifier and prover, which makes them susceptible to mordern security attacks. This paper presents SEDAT, a novel methodology for remote attestation of the device via a security enhanced communication channel. SEDAT performs hardware, firmware, and software attestation. SEDAT enhances the communication protocol security between verifier and prover by using the Single Packet Authorization (SPA) technique, which provides replay and Denial of Service (DoS) protection. SEDAT provides a way for verifier to get on-demand device integrity and authenticity status via a secure channel. It also enables the verifier to detect counterfeit hardware, change in firmware, and software code on the device. SEDAT validates the manufacturer's root CA certificate, platform certificate, endorsement certificate (EK), and attributes certificates to perform platform hardware attestation. SEDAT is the first known tool that represents firmware, and Integrity Measurement Authority (IMA) event logs in the Canonical Event Logs (CEL) format (recommended by Trusted Computing Group). SEDAT is the first implementation, to the best of our knowledge, that showcases end to end hardware, firmware, and software remote attestation using Trusted Platform Module (TPM2.0) which is resilient to DoS and replay attacks. SEDAT is the first remote verifier that is capable of retrieving a TPM2.0 quote from prover and validate it after regeneration, using a software TPM2.0 quote check. All source code, tools, and kernel patches are open-sourced via BSD 2-Clause License. 

\end{abstract}


%
\IEEEpeerreviewmaketitle

\section{Introduction}
\par Traditional computing and Embedded devices are proliferating into numerous and diverse aspects of everyday life. These devices are utilizing in different domains ranging from tiny personal gadgets to large industrial systems. the amount of such devices connected to the Internet is growing rapidly and securing these devices and our electronic infrastructure becomes increasingly difficult, in particular because a large fraction of devices cannot be managed by security professional nor can they be protected by firewalls. this devices are most likely to be susceptible to attacks like - hardware modification, or counterfeit can cause serious security challenges as e.g., see Chinese supply chain hardware attack \cite{Schneier2018}. Malware infestation involves modifying a device's firmware and software and replacing benign code with a malicious one. Which can destroy physical equipment (e.g., see Stuxnet \cite{Vijayan}) or enables a more sophisticated attack threatening the safety of the users (e.g., see Jeep hack \cite{Schneider}).  This increasing importance confronts developers with new challenges. One of them is the verification of the identity and integrity of a device by the trusted remote attestation tool called remote verifier. 
\begin{figure}[h]
\begin{center}
\includegraphics[width=3.5in]{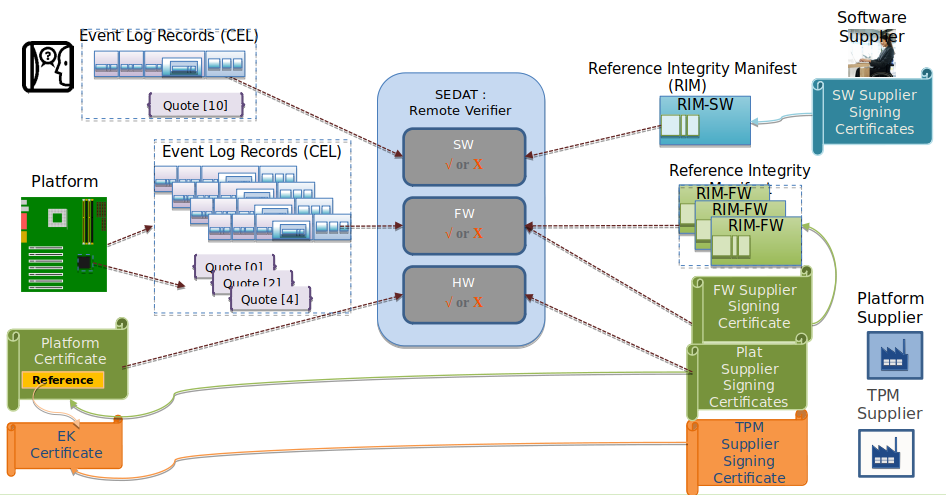}
\end{center}
\vspace{-2ex}
\caption{SEDAT: Remote Verifier
\vspace{-2ex}
\label{fig:SEDAT}}
\end{figure}    
\par Fig-1 shows high-level design of trusted remote verifier - Security Enhanced Device Attestation with TPM 2.0 (SEDAT) to attest untrusted devices. As can be seen, the verifier has to attest three components: hardware, firmware, and software to fully attest a device. Device suppliers will provide platform root certificates, endorsement certificates, platform attributes certificate. The device will have firmware and software modules that need to be unchanged and executed in the correct order to ensure integrity. This execution of the firmware and software module will create event records in form of firmware and IMA event logs, which will be explained in detail in the coming section. Fig-1 has Reference Integrity Manifest (RIM) for both firmware and software event logs, which will provide golden measurements based on Trusted Computing Group (TCG) 's new recommendations. Details of each module and working will be discussed in the following sections.

\par Remote Verifier (RV) has to validate the integrity and authenticity of hardware, firmware, and software state on the untrusted device against known good state for attesting the device. If any or all of the states fail, then the device attestation result is failed. RV should be able to perform attestation on-demand to ensure the correct state of remote prover. RV can perform a quote check with TPM2.0 to validate Ek and signing keys. 


{\bf {Goals and Contributions:}}

\par In this paper, we present SEDAT, the first proof of concept work for remote attestation, to ensure the integrity and authenticity of devices via a security-enhanced communication channel. Designing such a verifier is a challenging task.As there are many possibilities of malfunctions, like, their could be counterfeit hardware, modified software or malicious prover will triggering the attestation process. Therefore, we analyzed the requirements for designing a secure attestation protocol and depict how to address them with minimal features and assumptions. Further, we identified possible use-cases where SEDAT can be applied. Our work brings the following contributions:  
\begin{itemize}
\item {\bf{Client provisioning}:} Provided tool for client provisioning. it is the process that enables the platform owner to register the device with a remote verifier using a security-enhanced Single Packet Authentication (SPA) technique.
  \item {\bf{Platform attestation}:} Provide remote verifier for platform which attests platform root CA certificates, endorsement certificates, platform bindings.
  \item {\bf{CEL firmware event logs}:} Provide tools to convert and verify firmware event logs into canonical event log (CEL) structure. also, validated upstreamed kernel patches for crypto align firmware event logs.
\item {\bf{CEL IMA event logs}:} Provide tools and kernel patches for getting IMA event logs into userspace and converting it into CEL format.
\item {\bf{Quote check}:} Provide tools to check the quote at a remote verifier.
\item {\bf{Secure authentication}:} Provide tools for secure connection and authentication for verifier and protects it from replay and DoS type attacks. 
\end{itemize}
 
\par {\bf{Outline}:} This is not correct at this time --putting some place-holder flow. Outline, Section 2 reviews related work, adversary models and Replay, DoS attacks on attestation. Section 3 provides required preliminaries and notations, identifies
the minimal requirements for secure RV, and describes SEDAT. The prototype implementation of SEDAT has described in Section 4; its application to collective attestation is explained in Section 5. Next, the security of SEDAT has evaluated in Section 7, and the paper concludes in Section 8.
\section{Problem Description}
\par Remote attestation is an interactive process between a trusted remote verifier (denoted as RV) and a potentially untrusted remote device called prover (denoted as RP).It allows a trusted RV to capture the state of a potentially untrusted remote device.Essentially, RV measures and takes hash of the software running on the RP, tranfers it to it self and matches to the golden measurement to attest the device. Remote attestation can be performed by Hardware-only, software-only, or hybrid techniques. Each approach has its merits and demerits. Hardware or hybrid approach provides better security assurance as they use immutable hardware as a root of trust. 
\par One approach to achieve better security is to equip these devices with a root of trust, such as a Trusted Platform Module (TPM), a Trusted
Execution Environment (TEE), and Software Guard Extensions (SGX), and then have that root of trust attest the state of the device or computations made. Many devices have a Trusted Platform Module (TPM) that fulfills these tasks.
\par Remote device attestation (RV) can be used to establish static or dynamic root of trust in cyber-physical and industrial controls systems.It can be used as building block for other security services and primitives, such as device provisioning, firmware updates, kernel software patching.
\subsection{\bf{TPM based remote attestation}:}
\par As shown in Fig-1 , to attest a device RV needs to validate hardware, firmware and software all three components of the remote device.In last decade, researcher and industry has tried to solve the remote attestation problem and provided couple of solutions but non of it solves all three sametime. For example National Security Agency (NSA) has open-sourced tool called HIRS\cite{HIRS} for complite platform / hardware attestation. but it does not cover firmware or software attestation, moreover it works on centos 7 only and supports old tpm2-tools and tpm2-tss Intel's stack for TPM based device attestation.Second,The International Business Machines Corporation (IBM) has open-sourced their implementation called Attestation Client Server (IBM-ACS) \cite{Goldman}. IT does platform certificates, hardware and software attestation but it uses IBM's TSS and Tools to communicate with software TPM and its not complitely supported for hardware or firmware TPM. Also, it does not have support to verify firmware and software event logs in Cannonical Event logs (CEL) structure. Google has their open-sourced implementation for remote attestation called go-attestation\cite{weeks}, which does not do platform attestation and start at AK as root of trust. Acadamic implementation called keylime\cite{keylime} has support for multiple platforms and languages for attestation but it misses platform certificate validations, firmware and software even logs validations. all of the above solutions do not have support for replay or Denial of Service (DoS) attacks as they all work on https client-server protocols. 

\par This movitaved our research to first figure out what threat and adversary model are open to address, followed by looking at limitations of available attestation schemes.  

\subsection{\bf{Threat and Adversary Model}:}
\par Following the adversarial models from \cite{Ibrahim2}, SEDAT has classified attacks on remote attestation in three categories.
\begin{itemize}
\item {\bf{Communication Adversary}:} Adversary has complete control over all
communication channels. it can do eavesdropping and/or inject/modify packets, delay or drop packets. in case of DoS attack, adversery will flood the remote verifier by sending multiple provisioning request and eventually brings the verifier down.
\item {\bf{Software Adversary}:} Adversary can exploit software vulnerabilities
to infect prover or verifier, read its unprotected memory regions, manipulate
its software state or fake identity of prover. 
\item {\bf{Mobile Adversary}:} In addition to adversaies soft capabilities, this sophiesticated mobile adversaries are capable of erasing all traces of its previous presence on prover which makes it not detected by remote verifier. Executing such a sophisticated attack requires the knowledge of the exact execution time of attestation.  
\end{itemize}

\subsection{\bf{Limitations of Current Solutions}:}   
\par In recent research papers and in practice TPM has been used for hardware root of trust and remote attestation. where the integrity and authenticity of platform is relied on TPM based attestation key (AIK or AK). 
\begin{itemize}
\item {\bf{AK limitation}:} AK is taken as hardware root of trust anchor for remote attestation. problem of taking AK as only hardware root of trust is, generation of AK comes few step later in device manufacturing process. AK is derived from endorsement key (EK) and EK is provided from platform/hardware vendor by fusing private key and exposing associated public key as explained in next section. So, AK based attestation fails to detect malicious or counterfiet hardware, the TPM module or both. Also, some implementations uses software seed and AK key for attestation.
\item {\bf{Firmware eventlogs limitation}:} System records the firmware events extended into PCR's in form of firmware event logs. Verifying firmware event logs assures boot sequence and firmware code states integrity. However problem with current state of firmware event logs are they are not crypto aliened format and not avaliable in user space. They are only available in UEFI shell. So, it needs to be read from ACPI table to user space. second, problem is they do not have sequence numbers for the events. so, its hard to read the events and keep track of them from the binary blob when transfered to remote verifier. details of the firmware event log generation is explained in Theory and requirement section of SEDAT.  
\item {\bf{Software eventlogs limitation}:} Same as firmware event logs, IMA event logs also does not have sequence numbers. Second major problem is, since it is stored into small firmware memory which is not ment for storing incremental large IMA event logs. It needs the mechanisum to free up the space once the blob is read in user space.
\item {\bf{Quote check limitation}:} Some remote attestation implemetations have used quote generation with nonce and checks it at the provers, but it will be more valuable to check quote and get same PCR values at the RV to ensure that there was no tempering in middle. 
\item {\bf{Protection to known attacks}:} Some non TPM based attestation works have tried to enchance the security against replay and DoS attacks but majority of the attestation framework takes the https protocol as secure communication channel.   
\end{itemize}

\par Above all reasons combined motivated our research and eventually resulting in implemetation of SEDAT. SEDAT: the first remote attestation scheme, which performs hardware, firmware and software attestation and it is completely secure against Denial of Service (DoS), replay attacks.

\section{Theory, requirements, assumptions and limitations for New SEDAT}
\subsection{\bf{Theory and remote attestation requrements}:} SEDAT has identified remote device attestation problem needs to address five issues. the theory and requirement for each are explained in below sections, Assumptions and limitations considered while desinging SEDAT are discussed in next subsection. 

\subsubsection{\bf{Platform Validation}} It is common practice in industry that device vendor will not be the one and only vendor for all the hardware software modules compressed in the device. The device vendor will assemble the device by assembling all different parts in the manufacturing unit and send it to warehouse. After receiving the final product, the device vendor needs to validate the authenticity and integrity of all the modules present in the device. 
\begin{figure}[H]
\begin{center}
\includegraphics[width=3.5in]{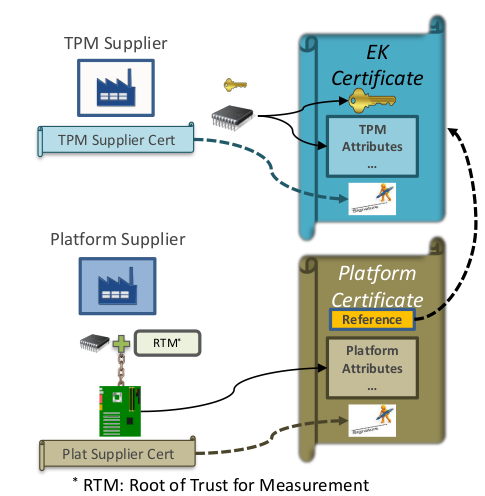}
\end{center}
\vspace{-2ex}
\caption{Platform and TPM supplier certificates binding, establishing platform ownership  
\vspace{-2ex}
\label{fig-2:binding}}
\end{figure} 

All hardware module manufacturers will provide modules' root CA certificate signed by the module manufacturer. The device vendor will validate all module certificates and binds them together with the platform and generates a self-signed platform certificate. The vendor creates a platform attributes certificate,   RV needs to verify hardware certificates and bindings. 
\begin{figure*}
\centering
\includegraphics[width=\textwidth]{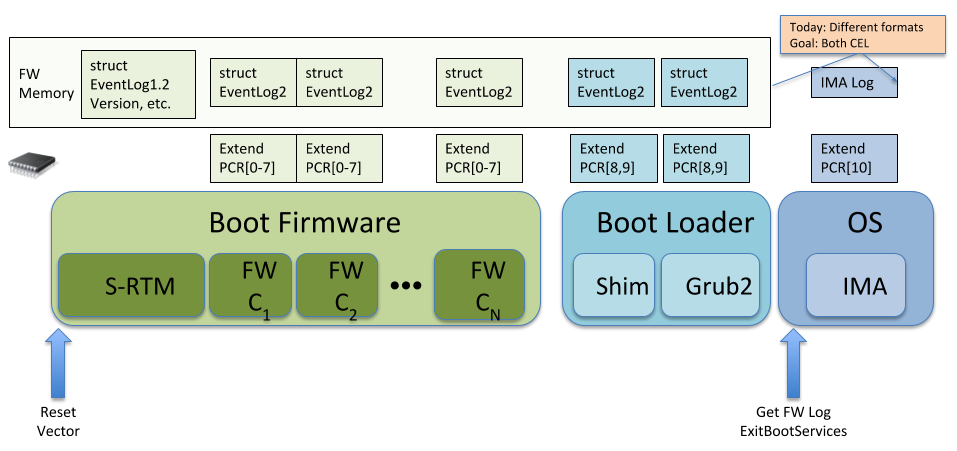}
\vspace{-2ex}
\caption{Boot Sequence of (x86 / UEFI / TPM2.0) 
\vspace{-2ex}
\label{fig-3:Boot Sequence}}
\end{figure*} 
\par Platform verification process has following steps, as depicted in Fig.2.
\begin{itemize}
\item {\bf{Step 1}:} TPM vendors will create endorsement private and public keys (EK) for the TPM platform. The private part of the endorsement key will be fused into the hardware, and the public key will be exposed for creating the platform attestation key (AK). This attestation key will act as a trust anchor in the hardware root of trust. EK is used for creating an endorsement certificate.
\item {\bf{Step 2}:} The platform supplier will provide a self-signed platform certificate. System user can create platform attributes certificate to get details about hardware modules are parts present on the device.
\item {\bf{Step 3}:} The platform supplier will reference  and bind the platform attributes and TPM certificate generated in the previous step. This step will make TPM to be exclusive to the platform. Platform attributes certificate will provide details about what other hardware modules are present onto the platform, which are mutable and non-mutable components.\item {\bf{Step 4}:} both EK and platform certificates are stored the in NV storage of TPM2.0.
\end{itemize} 

\par In order to attest platform RV needs to verify all of the above steps.

\subsubsection{\bf{Firmware Validation}} Firmware is a collection of codes stored on a small memory chip of a device. It provides the necessary instructions for the device to communicate with other hardware and software modules within and outside the device. The device uses flash ROM to store firmware, and it is semi-permanent unless it is changed or upgraded. Understanding the platform boot sequence (x86 / UEFI / TPM2) is a useful to perform firmware and software remote attestation. Fig.3 shows the boot sequence of X86 / UEFI / TPM2 based hardware device. 
\par Root of trust measurement will be done by following four basic operations namely \\ 
\begin{center} LOAD \end{center} \begin{center} MEASURE\end{center} \begin{center} EXTEND\end{center}  \begin{center} EXECUTE\end{center} 
\par  as seen from Fig-3, When the systems powers on, reset vector will first LOAD the Static Root of Trust Measurement (S-RTM) component of boot firmware. System MEASURES its code by taking a secure hash, EXTENDS it into PCR, creates first event record into firmware event logs in firmware memory. Next, the system EXECUTEs S-RTM module code and gives the information regarding the next boot image. The program counter will point to the next firmware image location, and the system will perform the same -  LOAD, MEASURE, EXTEND and EXECUTE operations on the next firmware component (e.g., FW C1). Each firmware component will follow the same boot sequence and will have an event extended into PCR0- PCR7. The system will generate an event record in firmware event logs. The last firmware boot component will point to the Shim or shimx64 as the next stage bootloader image. Which, in turn, will call grub or grub2, followed by loading Operating System (OS). These will have events extended in PCR8, and PCR9 and firmware event logs will have record of each event.
\begin{figure*}
\centering
\includegraphics[width=\textwidth]{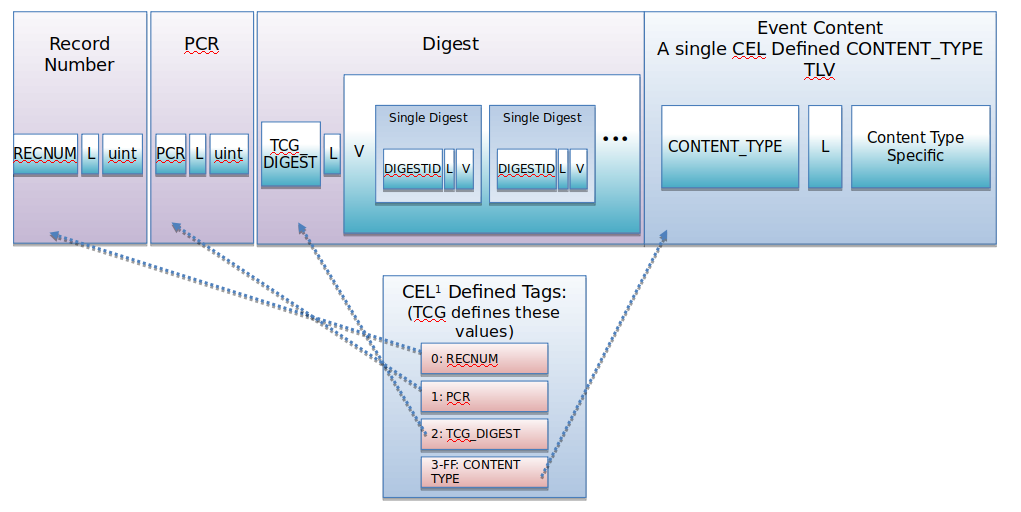} 
\vspace{-2ex}
\caption{CEL IMA eventlogs 
\vspace{-2ex}
\label{fig:ima eventlog}}
\end{figure*}  
\par As seen in Fig-3, today firmware and software event logs are in different format and does not have sequence numbers , which makes hard to transfer meaningful binary blob over to RV. one of the goal of SEDAT is to convert both event logs into Cannonical Event Logs (CEL) structure as recommended by TCG.

\par The malware's like ransomware tries to change firmware code or device boot sequence to lock the device and resources. So, the assurance of the integrity of firmware is important for device attestation along with the boot sequence integrity. RV needs to verify firmware event logs.  

\subsubsection{\bf{Software Validation}} Remote attestation of all the software on a device is a relatively hard and resource-heavy process. Instead, we can have a certain portion of the software modules attested, to ensure the critical portion of the software code and data is intact. Trusted Computing Group (TCG) has recommended a standard for software integrity check-called {Integrity Measurement Authority (IMA)}. Which signs, measures, and extends the IMA secure region of software into PCR10 of TPM2.0. It will also generate entry in IMA event logs. RV needs to verify IMA event logs for the IMA software integrity check.     
 
\subsubsection{\bf{Quote Validation}} TPM2.0 has a function called quote generation. The devices' AK, selected encryption algorithm, and PCR values are used to generate a quote. TPM2.0  uses nonce for adding freshness to the quote, which is sent from the verifier for the added layer of security and replay protection. RV needs to validate the quote.

The next section describes the assumptions and limitations we considered while designing SEDAT.

\subsection{\bf{Assumptions and Limitations of SEDAT}}  
\par SEDAT is implemented keeping hardware as root of trust using TPM2.0. All the devices provisioned with SEDAT requires to have hardware or firmware TPM2.0. Devices should have installed intel's latest TPM2-TSS, TPM2-abrmd and TPM2-tools. SEDAT assumes that, there is one time trusted secure channel for verifier to get all the root certificates from the device vendor. SEDAT uses single packet authorization for securing communication channel between prover and verifer. So, SEDAT assumes that pre-shared secret key is loaded at both ends before communication start between varifier and prover. It is protected from replay and DoS attacks but covering some advance attacks are out of scope of this research. All prover devices are required to be patched with provided IMA and firmware patches and latest linux kernel to get firmware and IMA event logs in CEL format. Protecting the prover or verifier from physical, side channel attacks is out of scope. SEDAT can be validated on Software TPM with intels TPM2 stack.
\section{Architecture building blocks for SEDAT}
\par After understanding theory and design requierments SEDAT is designed in moduler fashion.each module is performing dedicated task as follow.

\subsection{\bf{Device provisioner}} This is the first communication anchor between prover and verifier. Prover sends a hello message to the verifier. Verifier sends counter value to prover in response. Prover uses this counter value and to calculate Hashed One Time Password (HOTP) using pre-shared key and sends the HTOP to the verifier. The verifier compares the received HOTP with the one it has computed, if those matches, SEDAT enrolls the client to remote verifier by sending ack signal else the connection will is dropped. Prover sends its device information, Bios details, OS details to the verifier in the following message and device provisioning task is finished by recording response into database.
\subsection{\bf{Platform attestation}} This tool perform all the steps listed in platform validation subsection. also it take the ownership of the platform and TPM module as shown in Fig- 5. 
\begin{figure}[H]
\begin{center}
\includegraphics[width=3.5in]{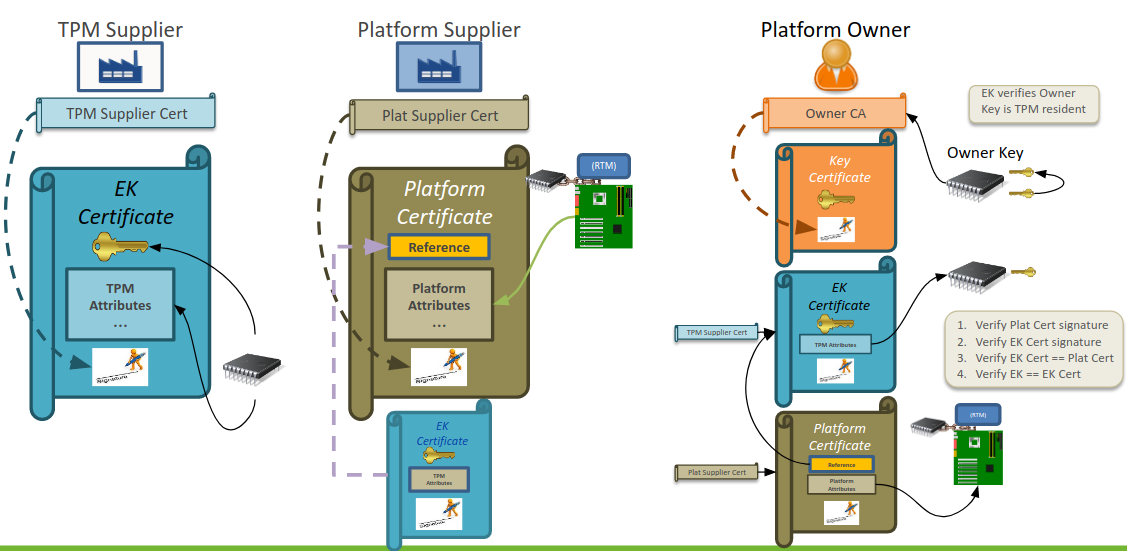}
\end{center}
\vspace{-2ex}
\caption{Platform and TPM supplier certificates binding, establishing platform ownership  
\vspace{-2ex}
\label{fig:binding}}
\end{figure}  
\subsection{\bf{Firmware and software event logs generation}} This are combination of patches and scripts which pulls the firmware and IMA event logs into user space and using tools it converts them into CEL format. \par Fig 4 depects the new Canonical Eventlog Structure (CEL) for IMA and Firmware events recommended by Trusted Computing Group (TCG). each field in the eventlog has Tag Lenght and Value (TLV) parameters. for firmware first event is TPM1.2 eventlog format as per pc client specification by TCG which has information about bios firmware versions, events are crypto aligned, supported hashing algorithms etc. second event onwards are actual events for firmware as explained before. it  will have PCR number which is 32 bit value between PCR0- PCR9 for firmware eventlogs and PCR10 for IMA. Digest will give information about hashing algorithm used for that event. Length of Digest field will be dependet on hashing algorithm used for that event. if it is SHA1 length will be 20 bytes, SHA256 length will be 32bytes  and so on.Value field in the Digest will hold extended PCR value. Event content field will hold the actual data in TLV format. on top of all this eventlog has the sequence number field for firmware and IMA eventlogs to make the records more meaningfull and easy to understand when transfer over to verifier.  
\subsection{\bf{Quote attestation}} Quote check is the mechanism used in TPM based attestation to validate identity and authentication of the platform. Mostly prover generates the quote and validates the quote locally. There is no known solution available, which does quote check at remote verifier. this tool is a quote verifier which runs couple of scripts on prover and gets quote at RV and following same process it regenerates the quote at RV and matches it. 
\par The next section explains workflow of the SEDAT framework.

\section{SEDAT : framework design}
\par The framwork of SEDAT works as follow
\begin{itemize}
\item {\bf{Client provisioning}:} In this step untrusted prover will establish secure communication channel with the remote verifier. 
\item {\bf{Get TPM certificate}:} Using TPM2.0 command get the TPM EK key and create endorsement certificate from the manufacturers root certificate site. 
\item {\bf{Get platform certificate}:} Using TPM2.-0 command to get platform certificate and run paccor to create platform attributes certificate. 
\item {\bf{Store certificates in TPM}:} Using TPM2.0 commands  store the platform certificate and endorsement certificate into NV storeage of TPM2.0.
\item {\bf{Take ownership}:} Using TPM2.0 command take ownership of the platform.
\item {\bf{Get eventlogs for Firmware and IMA in CEL format}:}  Using provided scripts get the firmware and IMA event logs from /sys/kernel/security/tpm0/binary* and /system/kernel/security/ima/binary* respectively and send it to RV. 
\item {\bf{Generate quote}:} Using tpm2\_quote command and added freshness nonce generate quote.  
\item {\bf{Validate quote at RV}:} based on received information at the remote verifier regenerate and check the  quote. this is implemented using IBM's software tpm as verifier needs not to have tpm module. 

\end{itemize}
\section{Implimentation/ Tools and techniques}  
\par Verifier and tools are implemented in go language and used postgres as backend database. we took insperation from National Security Authorities (NSA)'s tool for platform varification called HIRS. SEDAT verifier uses NSA's tool called paccor to generate platform certificate and platform attributes. it stores those two certificate along with manufacturer's root CA certificate into the postgres database as golden template. SEDAT transfers the golden CEL firmware and IMA event logs to the remote verifier using replay protected secure channel. we provide full control to verifier to enable or disable certificates, firmware event logs and IMA logs validation. 

\section{Usecases}
SEDAT can be applied to different application areas.here we showcase three usecases.
\begin{figure}[H]
\begin{center}
\includegraphics[width=3.5in]{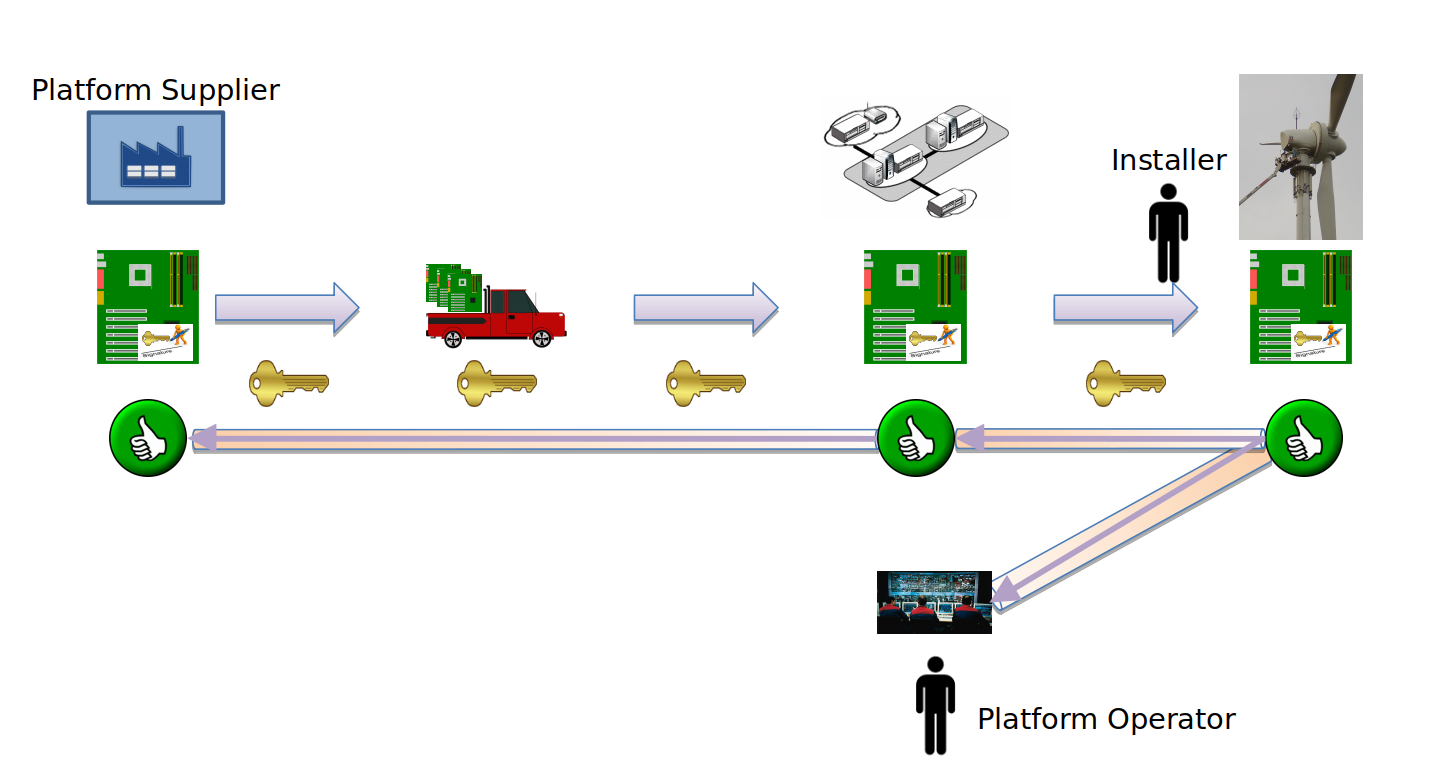}
\end{center}
\vspace{-2ex}
\caption{Supply Chain 
\vspace{-2ex}
\label{fig:Supply Chain}}
\end{figure}
\begin{itemize}
\item {\bf{Supply Chain Validation}:}
\par In supply chain platform supplier will be located in different geolocation and it transports the devices from the assembly line to warehouse, then it will go to retail location or to the installation facility. In this transport process, there are multiple untrusted anchors involved which can lead to counterfiet, substitute the devices.tpm based attestation provides the hardwae root of trust,signed key, certificate validation will increase security and trust. our verifier helps in tracking with reduced cost and increased trust. it also reduces the in situ installation and replacement cost.it is possible to remote key provisioning. key allows trusted remote configuration, trust channels using keys allows multiplexing connections reducing cabling costs.with SPA the communication is replay and DoS protected.\\ 
\begin{figure}[H]
\begin{center}
\includegraphics[width=3.5in]{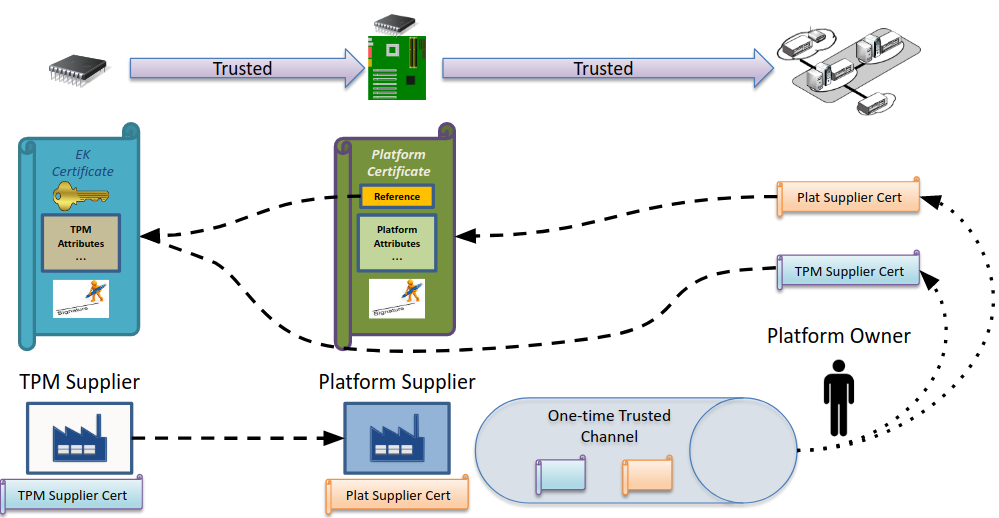}
\end{center}
\vspace{-2ex}
\caption{One-time secure channel 
\vspace{-2ex}
\label{fig: One time secure channel}}
\end{figure} 
 \par As shown in Fig.7 we need to have one-time trusted channel between platform owner and tpm and platform supplier to transfer securely the platform supplier cert and tpm supplier cert to platform owner and it will be binded and referenced as discussed before. we are using standard https connection for now. 
\begin{figure}[H]
\begin{center}
\includegraphics[width=3.5in]{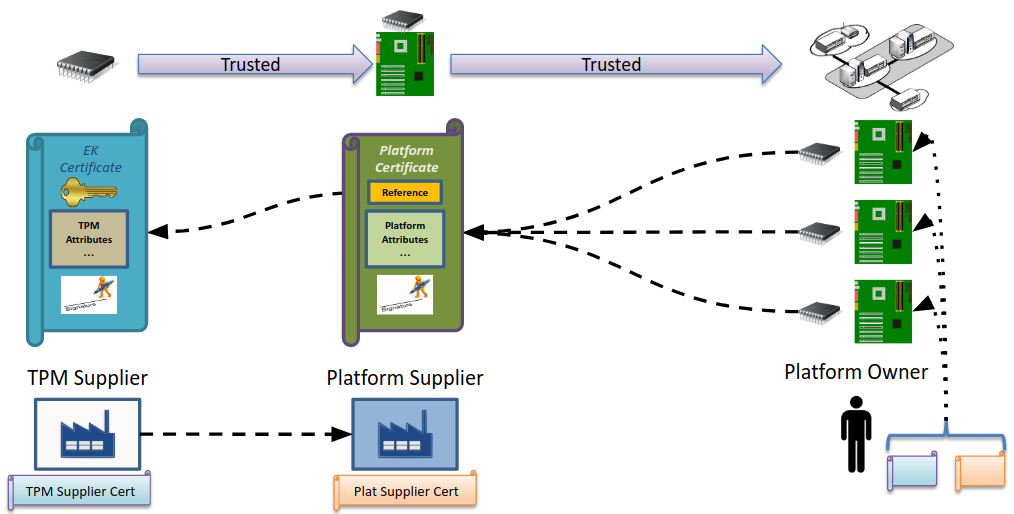}
\end{center}
\vspace{-2ex}
\caption{Supply Chain Validation
\vspace{-2ex}
\label{fig:Supply chain}}
\end{figure}
 \par As shown in Fig.8 shows supply chain validation uses case. the  connection for now. 
 \item {\bf{Inventory management}:}
 \par In large organization employer will give computers to emplyees which they use for all corporate works so integrity, authenticity and inventory management of those coputer devices are must. SEDAT can be used as remote attestation verifier as most of the computers now a days have TPM chip. SEDAT can also be laverage to list information of mutable and immutable components on computers as it was shown in \cite{HIRS}. this helps in boot time security and inventory management. if some employee has broken display or ethernet port while on vacation or travel and they replaced untrusted replacement part with SEDAT trust anchor installed verifier has golden state of all mutable immutable components on the device with was given to employee on day one. so next time when attestation takes place verifiers platform certs will not match and we can trigger an alert for appropriate action.
 \item {\bf{Industrial control systems}:}
 \par SEDAT can be used as remote verifier for controllers and embedded devices with hardware or firmware based TPM module. in those environment verifying the integrity and authenticity of the platform is key factor. 
 
 \end{itemize}

\section{Evaluation}
We have compared SEDAT with other solutions and found that SEDAT outperforms others by providing one stop solution for Hardware firmware and software remote verification  with single packet authentication to protect from replay and DoS attacks.below figure shows the evaluation report.\\
\begin{table}[H]
	\scalebox{0.92}[0.95]{
	\begin{tabular}{@{}lcccc@{}}
	\toprule
	\multicolumn{5}{c}{Comparision table - SEDAT v/s other solutions}                                                                                          \\ \midrule
	\multicolumn{1}{l|}{Parameters}              & \multicolumn{1}{c|}{IBM-ACA} & \multicolumn{1}{c|}{Sublime} & \multicolumn{1}{c|}{HIRS-NSA} & SEDAT \\ \midrule
	\multicolumn{1}{l|}{Endorsement certificate}         & \multicolumn{1}{c|}{Y}       & \multicolumn{1}{c|}{N}       & \multicolumn{1}{c|}{Y}        & Y    \\
	\multicolumn{1}{l|}{Platform certificate}            & \multicolumn{1}{c|}{N}       & \multicolumn{1}{c|}{N}       & \multicolumn{1}{c|}{Y}        & Y    \\
	\multicolumn{1}{l|}{Platform attributes Certificate} & \multicolumn{1}{c|}{N}       & \multicolumn{1}{c|}{N}       & \multicolumn{1}{c|}{Y}        & Y    \\
	\multicolumn{1}{l|}{Platform mutable components}     & \multicolumn{1}{c|}{N}       & \multicolumn{1}{c|}{N}       & \multicolumn{1}{c|}{Y}        & Y    \\
	\multicolumn{1}{l|}{CEL Firmware Event logs}           & \multicolumn{1}{c|}{N}       & \multicolumn{1}{c|}{N}       & \multicolumn{1}{c|}{N}        & Y    \\
	\multicolumn{1}{l|}{CEL IMA Event logs}                & \multicolumn{1}{c|}{N}       & \multicolumn{1}{c|}{N}       & \multicolumn{1}{c|}{N}        & Y    \\
	\multicolumn{1}{l|}{Quote generation}                & \multicolumn{1}{c|}{Y}       & \multicolumn{1}{c|}{Y}       & \multicolumn{1}{c|}{N}        & Y    \\
	\multicolumn{1}{l|}{Quote check at RA}               & \multicolumn{1}{c|}{N}       & \multicolumn{1}{c|}{N}       & \multicolumn{1}{c|}{N}        & Y    \\
	\multicolumn{1}{l|}{replay /DoS protection}      & \multicolumn{1}{c|}{N}       & \multicolumn{1}{c|}{N}       & \multicolumn{1}{c|}{N}        & Y    \\
	\multicolumn{1}{l|}{Multi-OS support}                & \multicolumn{1}{c|}{Y}       & \multicolumn{1}{c|}{Y}       & \multicolumn{1}{c|}{N}        & Y    \\ \bottomrule
	\end{tabular}
	}
\end{table}

\subsection{Future Work}
We are planing to enchance our varifier to include secure communication and replay protection.also we are interested in looking into secure communication protocol to transfer the manufacturers root CA and endorsement cert securely to the verifier as currently SEDAT assume that there is a one time secure channel for transfering those to verifier. we are motivated to close the loop of remote varifier and take actions once it detects a problem in attestation.

\section{Related work}
Remote Verifier (RV) allows a trusted entity to securely measure internal state of the remote unstrusted platform (prover). RV can be used to establish static or dynamic root of trust in cyber-physical and industrial controls systems.It can be used as building block for other security services and primitives, such as provisioning, updates, patches.Current attestation approaches fall into two domains namely collective attestation and single device attestation.
\begin{itemize}
\item {\bf{Collective Attestation}:} \par Traditional attestation schemes consider
only a single prover and verifier. Swarm/collective attestation aims
at scaling existing attestation schemes to networks of embedded
devices, by leveraging in-network verification \cite{Asokan}, and novel cryp-
tographic primitives \cite{Ambrosin}.
SEDA \cite{Asokan} investigates the security of swarms of embedded de
vices. It presents the first attestation protocol for large swarm,
allowing a central verifier to assess the trustworthiness of a million 
device swarms in order of seconds. It achieves this by distributing
the attestation burden across the swarm, allowing neighbors to
attest each other, and aggregating the attestation results in a hop-
by-hop manner. SANA \cite{Ambrosin} enables low verification overhead due to
the integration of a novel Optimistic Aggregate Signatures (OAS),
which is a generalization of aggregate and multi-signatures. Finally,
DARPA \cite{Ibrahim} aims at detecting software compromise and device
capture in embedded networks. Since DoS attacks on collective
attestation are more significant, as it allows to adversary to disturb
a large network by targeting one device, SANA \cite{Ambrosin} proposes using
secure tokens obtained from a trusted third party to prohibt scaling
DoS attacks to large networks. However, SANA uses expensive
public key cryptography which imposes additional overhead on the
Prv, and is vulnerable to DoS attacks based on the digital signature
verification procedure.
\item {\bf{Single Device Attestation}:} \par It has three main categories:Software-based attestation schemes \cite{Gardner,kennell,Seshadri,Seshadri2,Seshadri3,Seshadri4}. does not require secure hardware. It enables attestion of legacy and low-end embedded devices with some assumptions. These assumptions are: adversery is not active during the attestation process,the attestation code and implementation are optimal,and presence of an out-of-band authentication channel. Due to this reasons, security of software-based attestation schemes has been challenged by \cite{Castelluccia,Sankar1, Wurster} and their applicability and reliability was limited. Hardware-based attestation schemes \cite{Kovah, McCune, McCune2,Petroni ,Sailer, Schellekens} provide better security guarantees. Software/Hardware Co-design or Hybrid
schemes such as  \cite{Brasser, Eldefrawy, Francillon, Koeberl} provides examples of minimal hardware-based features required for enabling secure remote attestation. Such security features are as simple as a Read Only Memory (ROM), and a simple Memory Protection Unit (MPU).
 
\par SEDAT is the first remote attestation scheme, which performs hardware, firmware and software attestation and it is completely secure against Denial of Service (DoS), replay attacks.  
\end{itemize}

\section{Conclusion}
In this paper we have presented proof of concept work for remote verifier to perform end to end attestation of hardware, firware and software of untrusted device with tpm2.0. to the best of our knowledge SEDAT is the first solution to demornstrate one stop solution for all three components varification with INTEL's /TCG recommended tpm2-tools stack. we are the first one to auther tools for represnting IMA and Firmware eventlogs into CEL format. all codes is open-sourced and available for future research.


\section*{Acknowledgment}
\par
The authors would like to thank faculties of University of Maryland Baltimore County (UMBC) and research group at General Electric global research for funding, actively participating and guiding the research work.




\bibliographystyle{IEEEtran}
\bibliography{references}

%
\end{document}